\documentclass[11pt]{article}
\usepackage{graphicx} % Use pdf, png, jpg, or eps§ with pdflatex; use eps in DVI mode
\usepackage{color}
\usepackage{xcolor}
\usepackage{hyperref}
\usepackage{bookmark}
\bookmarksetup{depth=2}
\usepackage{amssymb}
\usepackage{amsmath}
\usepackage{mathtools} % more complete amsymb
\usepackage{float}
\usepackage{pgfplotstable} % build table from file content
\newlength\Myfigwd
\usepackage{cleveref}
\crefname{figure}{}{figures}
\creflabelformat{figure}{#2\txtkublue{\textsf{\textbf{\footnotesize Figure #1}}}#3}
\crefname{table}{}{tables}
\creflabelformat{table}{#2\txtkublue{\textsf{\textbf{\footnotesize Table #1}}}#3}
\crefname{section}{}{sections}
\creflabelformat{table}{#2\txtkublue{\textsf{\textbf{\footnotesize Section #1}}}#3}
\usepackage{booktabs}
\usepackage{array}
\usepackage{tocloft} % Provides control over the typography of the Table of Contents, List of Figures and List of Tables, and the ability to create new ‘List of ...’. The ToC \parskip may be changed.
\usepackage{colortbl}
\usepackage[stable]{footmisc}   % allow footnotes in section headings
\usepackage[normalem]{ulem}               % struck-out in mathmode
\usepackage{braket}
\usepackage{url}
\usepackage{upgreek}
\usepackage{bm}
\usepackage{fullpage}
\usepackage{cite}
\usepackage{subcaption}
\usepackage{outlines}
\usepackage{appendix}
\usepackage{chngcntr}
\usepackage{etoolbox}
\usepackage{xspace}
\usepackage{multirow}
\usepackage{authblk}            % affiliation
\usepackage{tikz}
\usepackage{units}              % \nicefrac{}{}
\usepackage{ifthen}             % allow conditionals in newcommands
\usepackage{pdfpages}           % add entire pdfpages

\numberwithin{equation}{section}
\numberwithin{figure}{section}

\newcommand\ignore[1]{} % ignore content

\newcommand{\be}{\begin{equation}}
\newcommand{\ee}{\end{equation}}
\newcommand{\bea}{\begin{eqnarray}}
\newcommand{\eea}{\end{eqnarray}}

\renewcommand{\vec}[1]{\ifthenelse{\equal{#1}{\ell}}{\boldsymbol{#1}}{\mathbf{#1}}}

\renewcommand{\epsilon}{\varepsilon}
\renewcommand{\phi}{\varphi}

\def\<{\langle}
\def\>{\rangle}

\def\1{\mathbb{1}}

\graphicspath{{graphics/}}

\title{Beam monitoring for radiotherapy from conventional to FLASH dose rates using Low Gain Avalanche Silicon detectors} 

\author[1]{Cristhian Cadena}
\author[1]{Antonio Cota Rodriguez}
\author[2]{Tonatiuh Garc\'{i}a Ch\'{a}vez}
\author[3]{Yao Hao}
\author[4]{Rachel Kovach-Fuentes}
\author[5]{Harold Li}
\author[4]{Wei Li}
\author[1]{Javier A. Murillo Quijada}
\author[2]{Saul Anibal Rodr\'{i}guez Ram\'{i}rez}
\author[2]{Christophe Royon}
\author[6]{Emil Schueler}
\author[6]{Brett D. Velasquez}
\author[4]{Pablo Yepes}

\affil[1]{Departamento de Investigaci\'{o}n en F\'{i}sica, Universidad de Sonora, Mexico}%
\affil[2]{Department of Physics and Astronomy, University of Kansas, Lawrence, KS 66045, USA}%
\affil[3]{Department of Radiation Oncology, Washington University School of Medicine, St. Louis, MO, USA}%
\affil[4]{Physics and Astronomy Department, Rice University, 6100 Main St., Houston, USA}
\affil[5]{Department of Radiation Oncology, University of Kansas Cancer Center, Kansas City, KS, USA}%
\affil[6]{Department of Radiation Physics, Division of Radiation Oncology, The University of Texas MD Anderson Cancer Center, USA}

\begin{document}

\maketitle

\begin{abstract}
We report the performance of low gain avalanche Silicon detectors (LGADs) for instantaneous electron and proton beam monitoring across dose rates ranging from conventional radiotherapy to the FLASH regime, benefiting from the fast response of these detectors of a few nanoseconds. The beam sources provide a dose rate greater than 40~Gy/s through pulses of widths 0.5, 1, 2 and 3~$\mu$s for electron beams and 3, 5, 10 $\mu$s for proton beams. Two different LGAD devices and silicon diodes are tested,
yielding a linear dose response for electron beams up to $\sim$450~Gy/s and for proton beams up to $\sim$12~Gy/s. Beyond the linear regime the response continues to increase with a reduced slope and no true signal plateau is observed, at least up to 1800 Gy/s for electrons and 150 Gy/s for protons. 
This study contributes towards the instantaneous monitoring of increasingly intense flash beams for radiotherapy using fast detectors such as LGADs since measurements can be performed every fraction of $\mu$s.
\end{abstract}

\section{Introduction}

External beam radiotherapy (EBRT), with electron and proton acceleration devices able to produce particle beams with energies in the MeV regime, have been used since the 1960's to treat cancer tumors~\cite{koka2022technological}. %Electron and proton beams are currently widely used for EBRT~\cite{koka2022technological}. 
FLASH radiotherapy (FLASH-RT) refers to particle beams with ultra-high dose rate delivery (UHDR), reaching mean dose rates greater than 40~Gy/s, up to 10,000's Gy/s, well above conventional radiotherapy procedures with usual rates $\sim$0.01--0.1~Gy/s~\cite{schueller2020uhdpulse}. This alternative treatment is currently under research for a wider use in clinical trials. %It can be used with either electron or proton beams. 
It was recently demonstrated that this new technique is effective to affect  cancer cells with substantially reduced toxicity in health tissue with respect to standard EBRT treatments with conventional dose rate ~\cite{favaudon2014flash}. FLASH-RT treatment was first tested on humans in 2019, using an electron beam at Lausanne University Hospital~\cite{bourhis2019firsthumanflash} and proton clinical trials have been recently performed over mice and humans using proton pencil beam scanning~\cite{cunningham2021protonflash, jin2023protonflashclinical}.

Treatment at such high rate encounters ionization chamber limitations to measure instantaneous and integrated doses due to the required higher time resolution and presence of saturation effects~\cite{Petersson2017, McManus2020, Romano2022}. Achieving a successful clinical beam pulse characterisation and monitoring from conventional to flash dose rates is necessary for precise dose application in future tumor EBRT treatments, especially at FLASH-RT rate. Precise instantaneous dose rate determination will also allow the development of appropriate radiation protection automated procedures.

Many developments and studies have occured recently related to flash beam therapy and dose measurements. For protons, the MoVe-IT group~\cite{Monaco2023} demonstrated Low Gain Avalanche Detectors (LGADs) counting therapeutic proton beams with less than 1\% inefficiency up to very high fluence rates, and a dedicated SiC characterisation for proton UHDR dosimetry has now also been published~\cite{LopezPaz2025}.

The goal of our paper is to test LGAD detectors, originally developed at CERN, in a practical clinical monitoring methodology. We perform online dose measurements at the $\mu$s timescale using two methods. The spike counting method provides beam diagnostics and pulse structure characterisation at low dose rates (up to $\sim$1--2~Gy/s for our sensor sizes~\cite{Yepes2026}. The signal area integration method is the primary dosimetry tool, extending operation to dose rates of several hundred Gy/s. Both lead to a per-pulse dose measurement, fundamental for clinical use and for developing automated mechanisms to interrupt the beam if the delivered dose deviates from the treatment plan.

LGADs~\cite{Pellegrini2014LGAD} are characterised by fast rising time responses of 20--50~ps~\cite{Agapopoulou2020, MicronLGAD2025} and an internal gain layer, making them well suited for beam monitoring in EBRT. Their nanosecond-scale signal duration~\cite{petersson2017high, mcmanus2020challenge, romano2022ultra} allows individual particles to be resolved at moderate dose rates. LGAD performance has been validated against standard ionising chamber measurements for conventional electron beams~\cite{isidori2021lgad} and for 60~MeV proton beams up to $\sim$7.5~Gy/s~\cite{Bellora2025LGADProtonTherapy}. Recent tests of semiconductor-based devices demonstrate that the precision of dose measurements depends critically on both the data processing methodology and the electronic design~\cite{isidori2021lgad, Knopf2023LGADAlpha}.

In this work, we investigate the response of LGAD devices and silicon diodes to 7~MeV electron beams and 70~MeV proton beams from clinical accelerators at the MD Anderson Cancer Center and the Siteman Cancer Center. The detector response to conventional and FLASH-RT dose rates has been studied in detail.

The paper is organised as follows. Section~2 describes the detectors, the laser characterisation setup, the beam test setup, and the two dose measurement methods. Section~3 presents the results: the laser sensor scan, the electron beam measurements, and the proton beam measurements. Section~4 summarises the findings and outlines future directions.

\section{Materials and Methods}

\subsection{Silicon detectors}

The first LGAD Si sensor used in the tests, called LGAD 1, was provided by the research teams at the University of Kansas (KU) and The European Center for Nuclear Research (CERN). The second sensor, called LGAD 2, was provided by the Brookhaven National Laboratory (BNL). An additional setup including a single Si diode incorporated to the BNL device was also tested for reference.

The LGAD 1 silicon sensor of dimension 2.9$\times$0.5 mm$^2$ is 
described in Ref.~\cite{Albrow2014}. 
The readout board hosting the sensor and the amplification chain, developed at the University of Kansas, comprises two-stage trans-impedance amplifiers and a 20$\times$20mm$^2$ pad providing bias to the Si sensor to a maximum value of $\sim$500 V.  
Detailed features are described in Ref.~\cite{isidori2021lgad}, where the results using a similar detector to characterize an Elekta linac beam at the University of Dublin are also described.

\begin{figure}
\centering 
\includegraphics[width=0.4\textwidth]{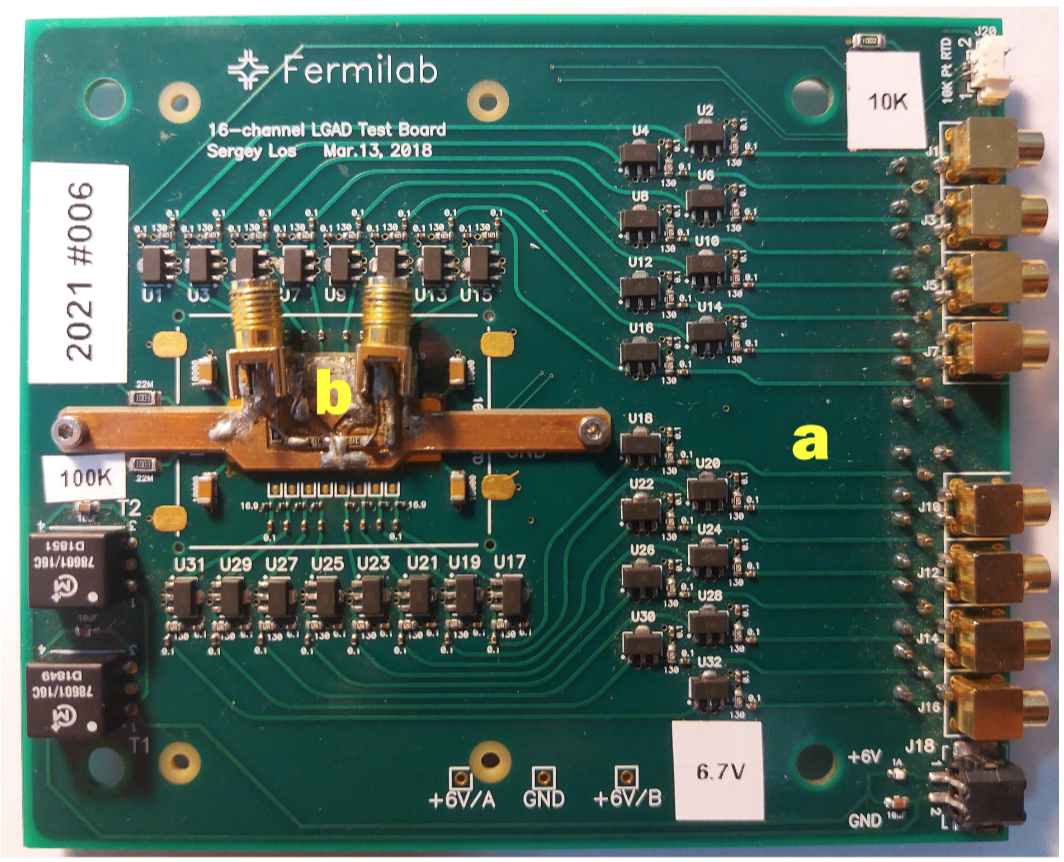}
\includegraphics[width=0.55\textwidth]{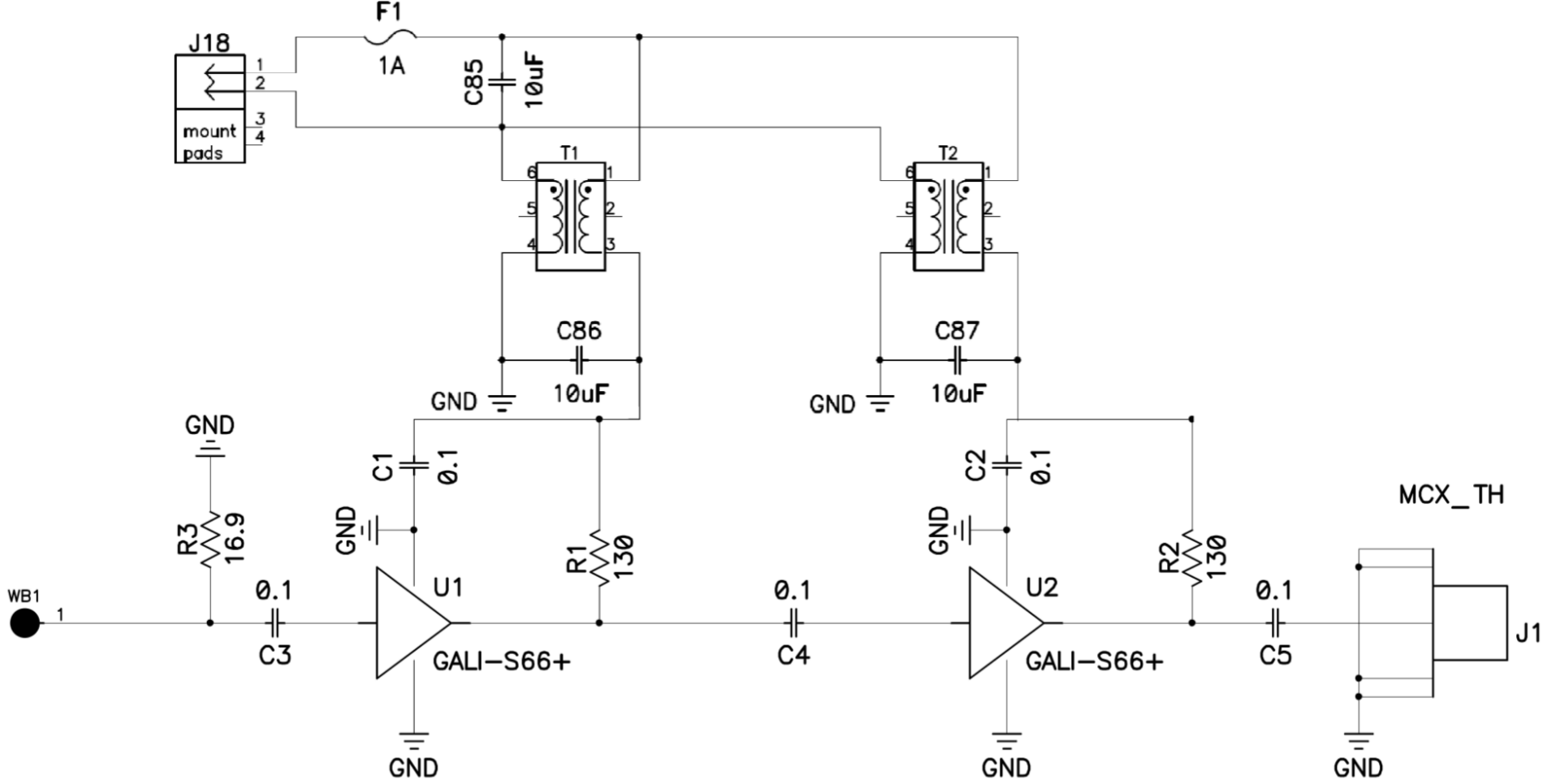}
\vspace{0.4cm}
\includegraphics[height=0.36\textwidth]{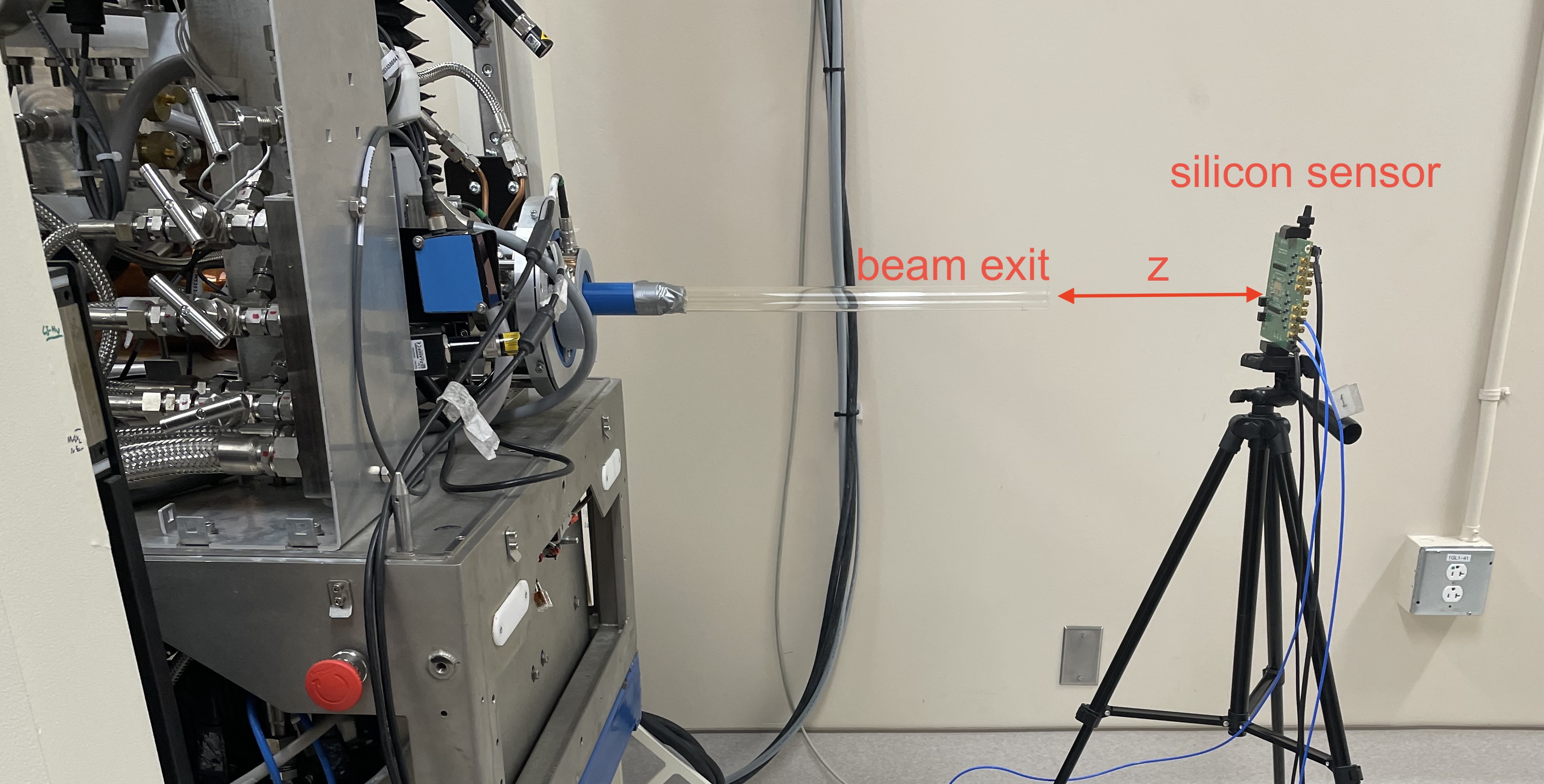}
%
%\includegraphics[height=0.50\textwidth]{images/KU-varian.jpeg}
%\hspace{0.3cm}
%\includegraphics[height=0.50\textwidth]{images/IMG_8854.jpeg}\\
\caption{\label{fig_BNL-sensor-board}Top: LGAD 2 and silicon diode board device configuration with a two stage amplifier system (top left) and detailed electronics diagram (top right).
Bottom: Experimental setup showing the beam source directed towards the silicon device connected to the oscilloscope. }
\end{figure}

The LGAD 2 device is a silicon sensor with an active transverse area of 1.3$\times$1.3 mm$^2$ and a depth value of 30$\mu$m~\cite{rios2025aclgads}. An additional setup with a single Si diode is also used for comparison. The amplifier board is a fast two stage amplifier relying on mini-circuits with low power GALI-S66+ chips as indicated in Fig.~\ref{fig_BNL-sensor-board}, top. It operates at a 1 GHz bandwidth with a gain value of $\sim$ 35dB with reported resolution of $\sim$2 ps. The LGAD mounting pad can be seen at the centre of the photo at the top of  Fig.~\ref{fig_BNL-sensor-board} with signal wire bond pads located at the top.

Both LGAD readout boards were originally developed for timing applications at the Large Hadron Collider, where they represent the state of the art in fast silicon detector readout. Their use constitutes the first systematic test of such detector and readout boards in a medical FLASH-RT context. While their nanosecond-scale timing performance is ideally suited to the spike counting and area integration methods described in this paper, their bandwidth and dynamic range were not optimised for the high instantaneous currents encountered at the highest flash dose rates, and this is reflected in the non-linear response discussed in Section 4.

\subsection{Beam test setup}

Our goal is to test the LGAD silicon detectors and to determine up to which dose rates we can measure the dose instantaneously (every $\sim \mu$s). We use a set of electron and proton beams with the possibility to obtain doses from conventional to FLASH-RT regime.

As indicated in  Fig.~\ref{fig_BNL-sensor-board}, bottom, the general strategy to probe the silicon device consisted in testing it using electron and proton beams, along the longitudinal beam direction, at a given distance $z$ from the beam exit. The silicon device was connected to an oscilloscope that recorded the voltage profiles as  function of time. In addition, different values of low and high voltages were applied to the silicon detector.

As an example, flash and conventional electron and proton beam sources located at the MD Anderson Cancer
Center in Houston, Texas and Siteman Cancer Center in St. Louis Missouri were used. 
An Oriatron eRT6 device providing conventional and FLASH-RT electron beam rates~\cite{Jaccard2018Oriatron} was used together with 
a polycrystalline collimator at the beam exit pointing towards the silicon detector to enhance the beam collimation. The exposed devices were located at different $z$ distances (0-270 cm) from the collimator position at the beam exit. A laser device was used to ensure proper alignment between the beam direction and the sensor location. Reference doses from conventional ionization chamber devices were also recorded. 
In addition, in order to test the response to proton pulses with ultra high doses,  a Mevion s250-fit accelerator providing FLASH proton beams at the Siteman centre in St Louis was also used.

Within a distance of $\sim$2 m, the dose rates originating from the electron beam, as measured using a clinical charge integration dosimeter, decrease by over two orders of magnitude. Depending on the voltage applied to the Oriatron machine, doses can go from $\sim$50 to  3$\times$10$^4$ Gy/s as a function of distance or to rates up to 2$\times$10$^5$ Gy/s and 3$\times$10$^5$ Gy/s. For the electron beam measurements with LGAD~1, the beam was deflected towards the detector using a $\sim$1~T magnet, in order to suppress the $\gamma$ background that is co-produced by the Oriatron and would otherwise contaminate the signal. 

Throughout this paper it is important to distinguish between two related but distinct quantities. The \textit{instantaneous dose rate} is the dose rate delivered during a single beam pulse, which can reach values of $10^3$--$10^5$~Gy/s close to the beam exit. The \textit{mean dose rate} is averaged over a full treatment session including the inter-pulse gaps, and is typically much lower; it is the mean dose rate that determines whether a treatment falls in the FLASH regime ($>$40~Gy/s). A detailed discussion of dose rate definitions and their implications for UHDR dosimetry can be found in Ref.~\cite{Yepes2026}.

It is important to note that the measurements reported throughout this paper characterise the response of the full acquisition chain: the silicon sensor, the front-end readout electronics, and the oscilloscope. For instance, the LGAD~2 board uses a two-stage GALI amplifier chain operating at approximately 1~GHz bandwidth. Any departure from linearity observed at high dose rates could therefore originate in the readout electronics rather than in the sensor itself, and this distinction will be discussed in the results section.

\section{\label{sec:level1}Methods to measure instantaneous doses}

Two different methods to measure doses instantaneously (every $\mu$s) were developed using the LGAD detectors 1 and 2 in an electron Oriatron beam. The first method implemented a spike counting procedure (each particle crossing our detector leads to a measured spike). The second method, described in the next section, is based on modifying the simple counting method by considering the measurement of the deposited charge in the detector via an integration of the signal area.

\subsection{The spike counting method}

The simplest method to measure doses instantaneously is to count the number of particles crossing the LGAD detector. This is possible thanks to the short response time of the detector, of the order of a few nanoseconds: each particle produces an identifiable spike in the voltage-versus-time waveform, and a threshold crossing count gives the number of particles directly. It is thus expected that this simple method will work if one particle enters every few nanoseconds.

\begin{figure}
\centering
\includegraphics[height=0.89\textwidth]{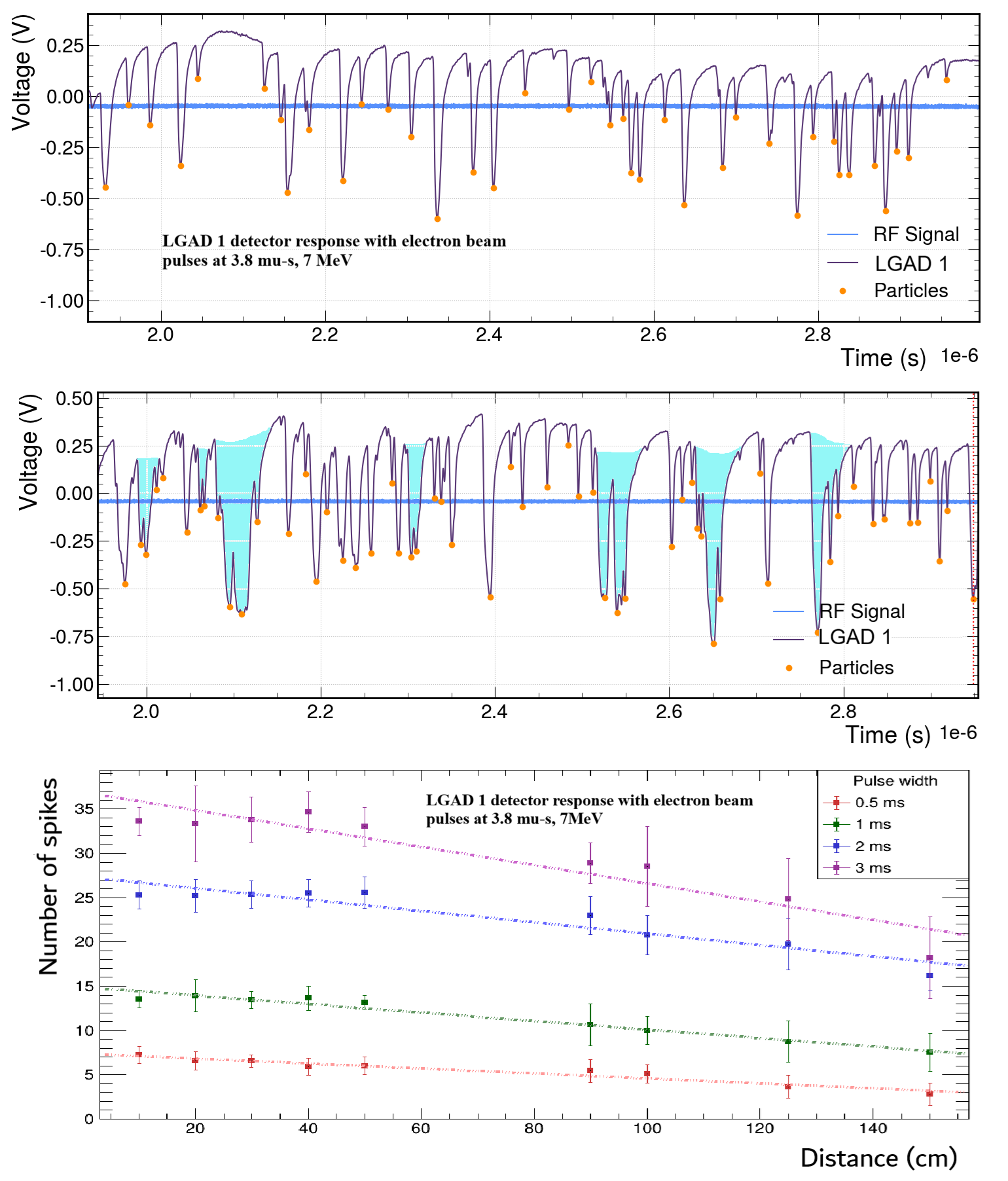}
\caption{\label{fig1b} Top: Example of readout from one bunch of electrons using LGAD 1. Center: Limit of the spike counting method because of the overlap between spikes. 
Bottom: Average spike counting result as a function of dose per pulse with uncertainty for different pulse width configurations and distances from 10 to 150 cm from the electron beam exit. The fitted linear fits give the following slope values for 0.5$\mu s$: 7.38 $\pm$ 0.235, 1$\mu s$:
14.92   $\pm$ 0.29, for 2$\mu s$:
27.34   $\pm$ 0.70, for 3$\mu s$:
36.92   $\pm$ 1.20, which are compatible with expectations (the delivered dose is in first approximation proportional to the bunch length)}
\end{figure}

Fig.~\ref{fig1b} illustrates the spike counting method in practice. The top panel shows the LGAD~1 signal as a function of time for a standard clinical pulse of 3.8~$\mu$s, where individual spikes corresponding to electrons entering the detector are clearly visible. The middle panel shows the onset of spike overlap at higher doses, confirming that the counting method reaches its ceiling at dose rates consistent with the $\sim$1--2~Gy/s estimate for our sensor area~\cite{Yepes2026}. This places the spike counting method firmly in the beam diagnostics regime: it is useful for verifying pulse structure and beam presence at low dose rates, but is not applicable for dosimetry at conventional or FLASH treatment rates. The bottom panel shows the spike counting result as a function of dose per pulse for different pulse widths and distances ranging from 10 to 150 cm from the beam exit: the count scales linearly with pulse duration as expected, confirming the method is working correctly in its valid operating range. We use a plastic shielding and the beam is deviated towards the sensor using a magnet of about 1T.   When the instantaneous dose (or the dose per pulse) increases, the number of particles entering the detector within few nanoseconds can be larger than one. The spike width is then larger, corresponding to multi-particle signal. It is then obvious that the simple counting will have to be modified by taking into account the signal area, or in other words by reweighting the simple counting method taking into account the signal area. This will be discussed in the next section. 

The spike particle counting method is proportional to the pulse width as shown in Fig.~\ref{fig1b}, bottom. The number of spikes for different pulse widths and distances is displayed as a function of the distance from the beam exit. We use a plastic shielding and the beam is deviated towards the sensor using a magnet of about 1T. Fig.~\ref{fig1b}, bottom, shows consistent spike counting rates with different pulse width variations for distances ranging from 10 to 150 cm from the beam exit. The fitted linear shapes is consistent with expectations since the number of spikes - or the dose measured by LGAD - is proportional to the pulse duration. Some discrepancy is observed for a pulse width value of 3 $\mu$s. The reason is likely due to the fact that this specific pulse width is close to the machine limitation.

The limits of this simple counting method can be estimated from basic scaling arguments~\cite{Yepes2026}. Assuming the optimistic limit of 1 particle resolved per nanosecond, and using the energy deposited by a 7~MeV electron in silicon, the maximum countable dose rate scales inversely with sensor area:
\begin{equation}
\dot{D}_{\max} \propto \frac{E_{\mathrm{dep}} \times (1~\mathrm{particle/ns})}{\mathrm{Area}}
\end{equation}
For LGAD~1 (active area $\sim$1.45~mm$^2$) and LGAD~2 (active area $\sim$1.69~mm$^2$), this yields a counting-mode ceiling of approximately 1--2~Gy/s. This is below the conventional radiotherapy rate and far below the FLASH threshold. The spike counting method is therefore not a dosimetry tool for clinical or FLASH-rate operation. Its value lies in \textit{beam diagnostics}: verifying the beam is on, characterising the pulse structure at low dose rates, and providing a qualitative check of beam quality. For dosimetry at clinically relevant dose rates, the area integration method described below is required.

\begin{figure}[h!]
\centering 
\includegraphics[height=0.50\textwidth]{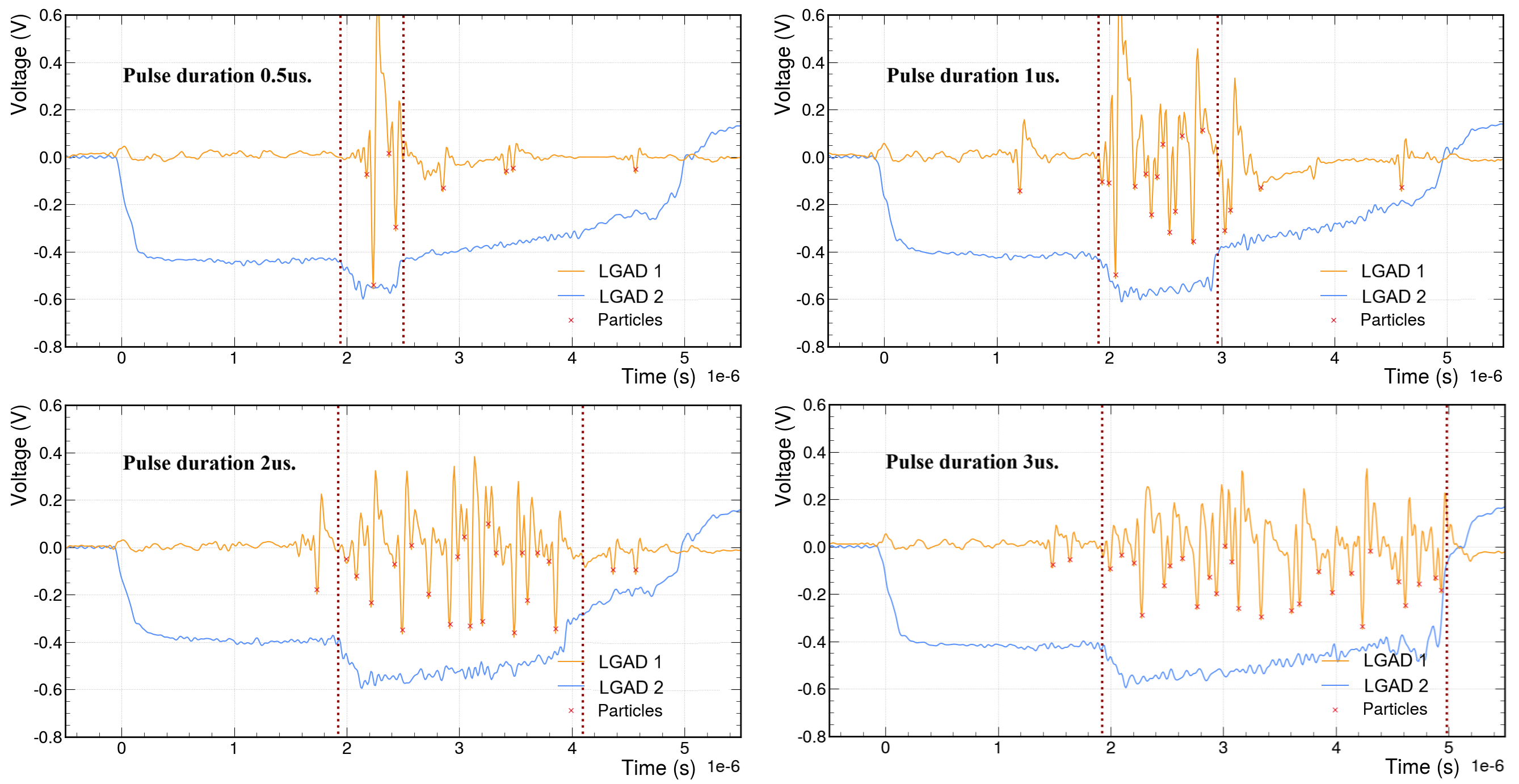}
\caption{\label{fig4b} Correlations between LGAD 1 and 2 detectors using electron beams. From top to bottom the pulse width is 0.5~$\mu$s, 1~$\mu$s, 2~$\mu$s and 3~$\mu$s respectively. A magnet was placed to bend the electron beam in order to clean the signal from $\gamma$ background. The devices were located at 0.5~m from the beam exit and 24~cm off the beam axis.}
\end{figure}

The results using LGADs 1 and 2 together are shown in Fig.~\ref{fig4b} for different pulse widths. LGAD~2 shows a well defined rectangular voltage step coinciding with the pulse duration, a consequence of its different readout electronics. A good coincidence and correlation between both detector signals is observed.

\subsection{Area integration and modified spike counting method}

At high instantaneous doses the spike counting method is strongly limited by overlapping spikes. We therefore developed a complementary method based on integrating the signal area over the beam pulse window, which is equivalent to measuring the total deposited charge in the detector. This method can be implemented in two ways, both referred to as the area method in the following.

The first approach uses a fixed integration range equal to the nominal pulse width. The integration window is centred on the signal maximum but is not symmetric around it.

\begin{figure*}
\centering
% --- Top Row ---
%\includegraphics[width=0.49\textwidth]{overlaid_Z20p0cm_HV7V_PW1p000us_BEAMElectrons_85V_scope-results-2025-10-15-0745.pdf}
\includegraphics[width=0.49\textwidth]{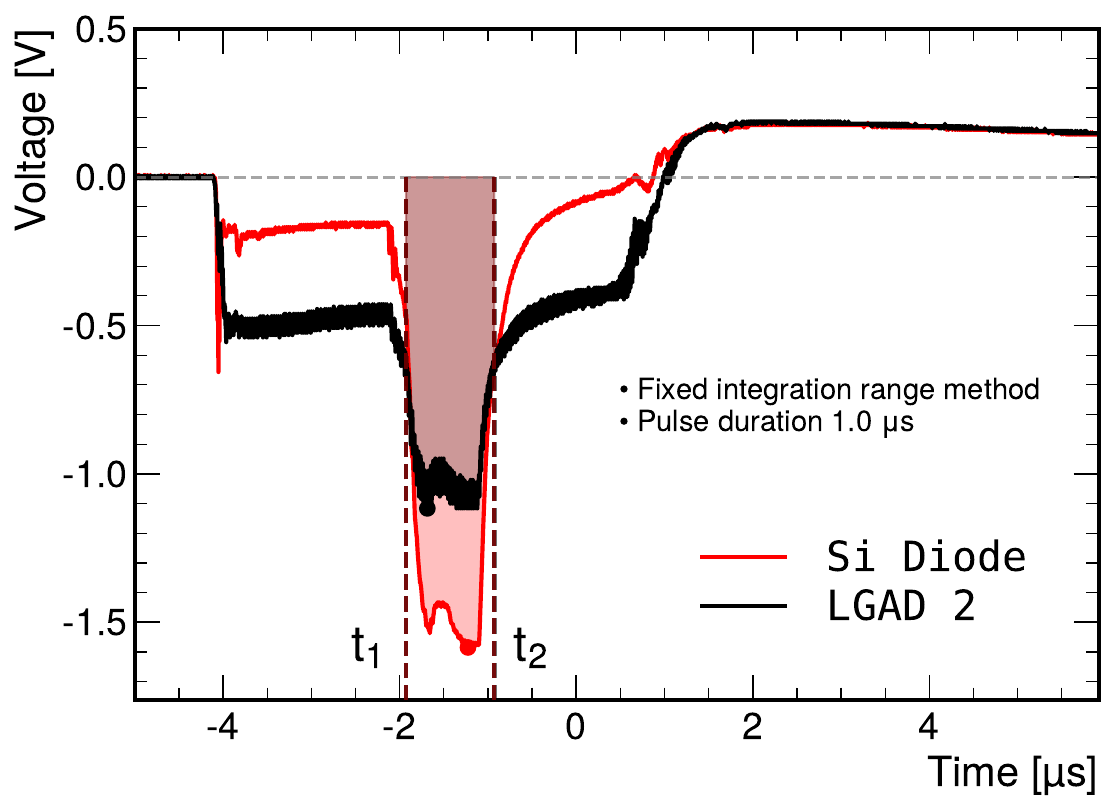}
\hfill  % Pushes the top-right image to the right margin
\includegraphics[width=0.49\textwidth]{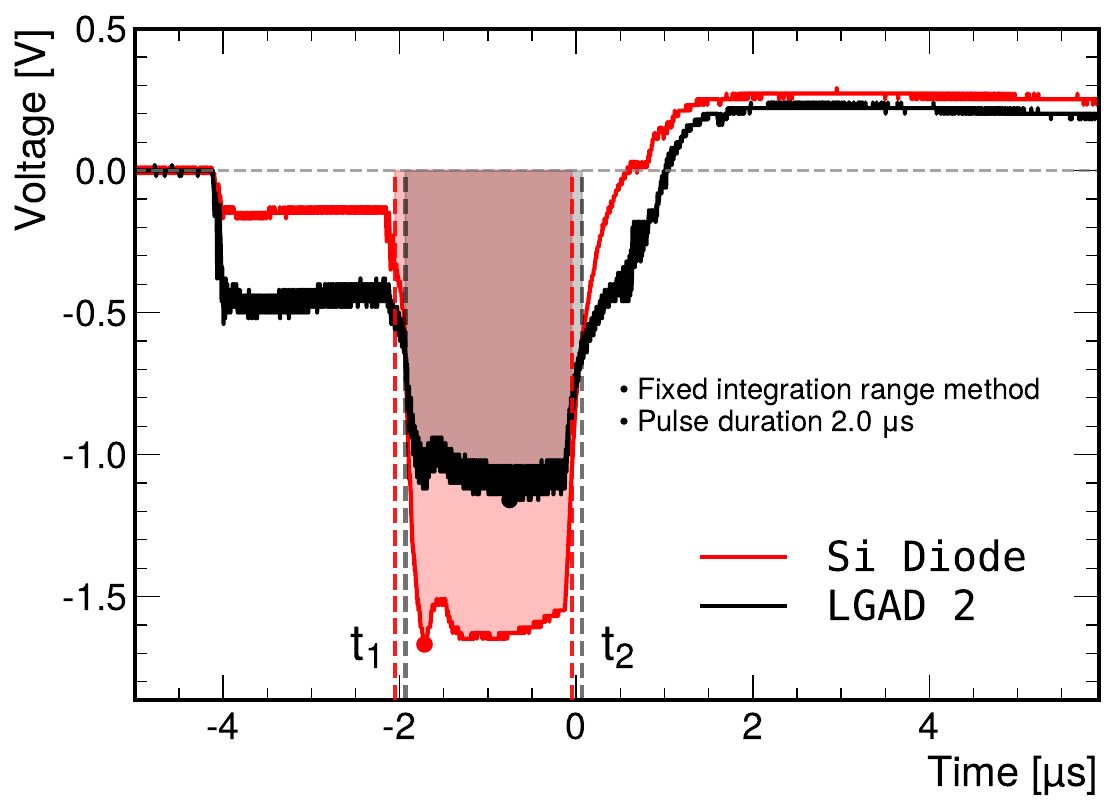}

\vspace{0.4cm} % Adjusts vertical spacing between rows cleanly

% --- Bottom Row ---
\includegraphics[width=0.48\textwidth]{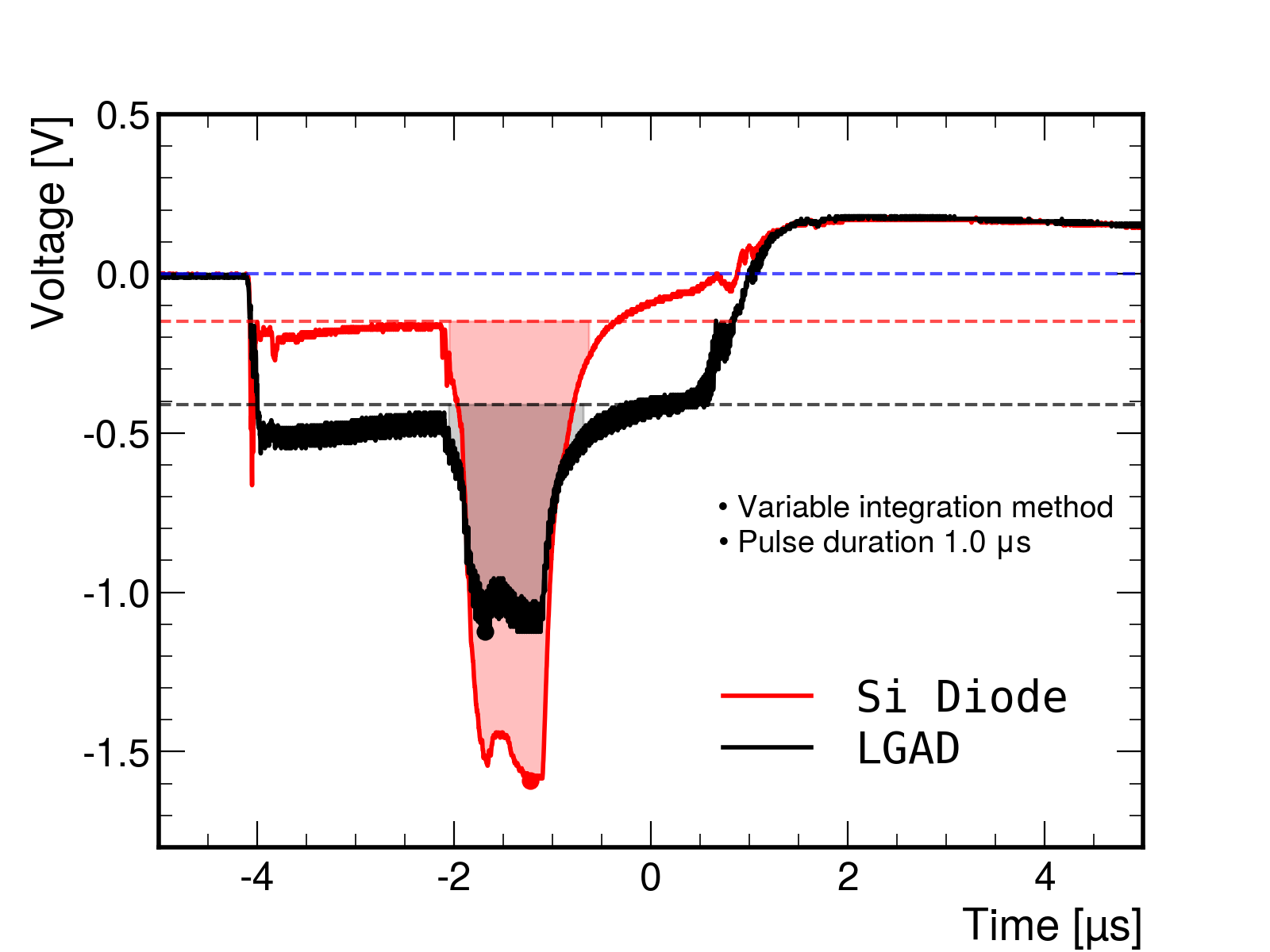}
\hfill  % Pushes the bottom-right image to the right margin
\includegraphics[width=0.48\textwidth]{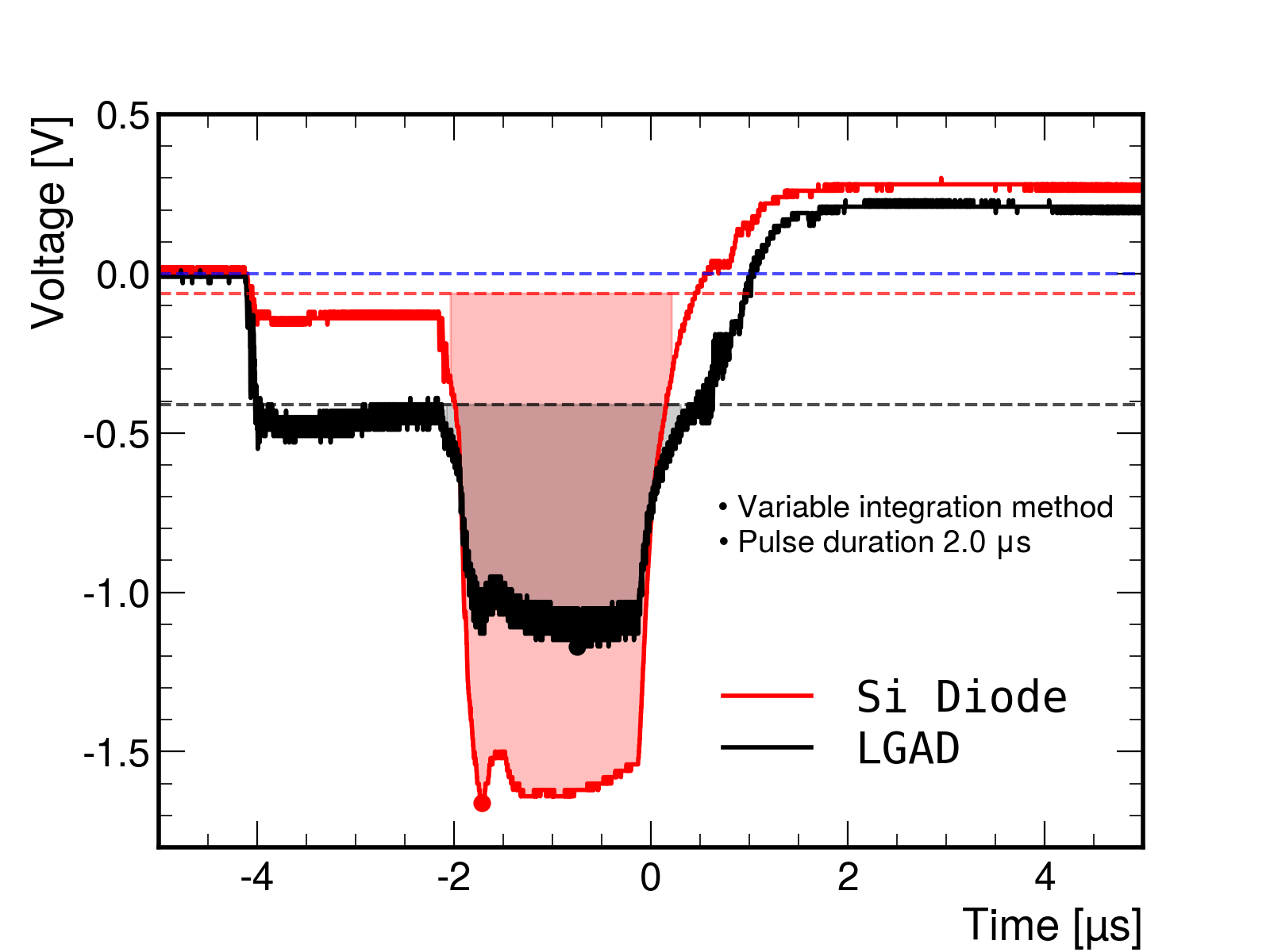}

\caption{Baseline-corrected waveforms using the full integration method (top) and the dynamic baseline method (bottom), obtained with LGAD~2 and the silicon diode at HV~=~7~V, $z=20$~cm, for pulse durations of 1~$\mu$s (left) and 2~$\mu$s (right). Note the negative baseline offset visible in both the LGAD~2 and silicon diode signals before and after the pulse, which is not present in the LGAD~1 measurements shown in Fig.~\ref{fig1b}. This offset is accounted for by the dynamic baseline method (bottom row).} 
\label{fig:waveforms_ch1_ch2_pulsewidth} 
\end{figure*}

The second approach determines the integration range dynamically without any a priori assumption about the pulse width. A sliding window scans the waveform from left to right and triggers at the second steep voltage drop, marking the end of the signal region. The algorithm then steps back to the local maximum to mark the start. The baseline is also determined dynamically by averaging the waveform immediately to the left and right of the signal region, which may result in a baseline below zero. This method removes the need to fix the signal width or baseline in advance, and as shown in Figure \ref{fig:waveforms_ch1_ch2_pulsewidth}, the resulting areas from this method are smaller than those from the fixed integration method due to the different baseline determination. Cases where no significant drop point was detected (usually due to very large noise) were discarded and not integrated by this method.

Both approaches yield consistent results; their differences are marginal in practice. Both are illustrated and validated against the reference ion chamber in Section 4.

Fig.~\ref{fig:waveforms_ch1_ch2_pulsewidth} shows example waveforms from LGAD~2 and the silicon diode at $z=20$~cm and HV~=~7~V for pulse durations of 1 and 2~$\mu$s. The diode produces a larger signal amplitude and a larger signal-to-background ratio. The top row shows the fixed integration range method. The points corresponding to the maximum of the signal are indicated in red and black dotted lines for both sensor signals. The integration range is fixed to the pulse width but is not symmetric around the  maximum.  The bottom row shows the dynamic baseline method.   The shaded areas indicate the integrated charge in each case, and the red and black dashed lines the dynamically determined baseline. The blue dashed line indicates 0V. The differences between the two methods are marginal.

We will now discuss the performance of the area integration method using electron and proton flash beams.

A notable feature of the LGAD~2 waveforms in Fig.~\ref{fig:waveforms_ch1_ch2_pulsewidth} is the negative baseline offset visible in the signal outside the pulse window, which is absent in the LGAD~1 waveforms (Fig.~\ref{fig1b}). The origin of this offset is not fully understood. RF pickup from the Oriatron electron machine is a plausible cause: the Oriatron delivers its macropulses as a train of picosecond-scale RF micropulses at 2856~MHz~\cite{Yepes2026}, and the LGAD~2 readout board -- with its 1~GHz bandwidth GALI amplifier chain -- may be more susceptible to this RF microstructure than the LGAD~1 trans-impedance amplifier board. In support of this interpretation, when LGAD~2 was operated in a synchrotron proton beam (PTC1, MD Anderson Cancer Center), whose beam structure consists of long spills without the high-frequency RF micropulse train characteristic of the Oriatron, no baseline offset was observed. This beam-dependent behaviour suggests that the effect is related to the specific RF time structure of the electron machine rather than being an intrinsic property of the LGAD~2 readout electronics. A systematic study with controlled RF shielding conditions is needed to confirm this hypothesis. Importantly, the dynamic baseline integration method described in Section~2 explicitly measures and subtracts the local baseline from regions immediately outside the signal window, so this offset is fully corrected for in the dose measurements and does not affect the results.

\section{Performance of the area integration method in beam tests}

In this section, we will discuss the performance of the area method (or the modified improved spike counting method) to measure instantaneous doses.

\subsection{Electron beams}

\begin{figure*}
    \centering
   \includegraphics[width=0.47\textwidth]{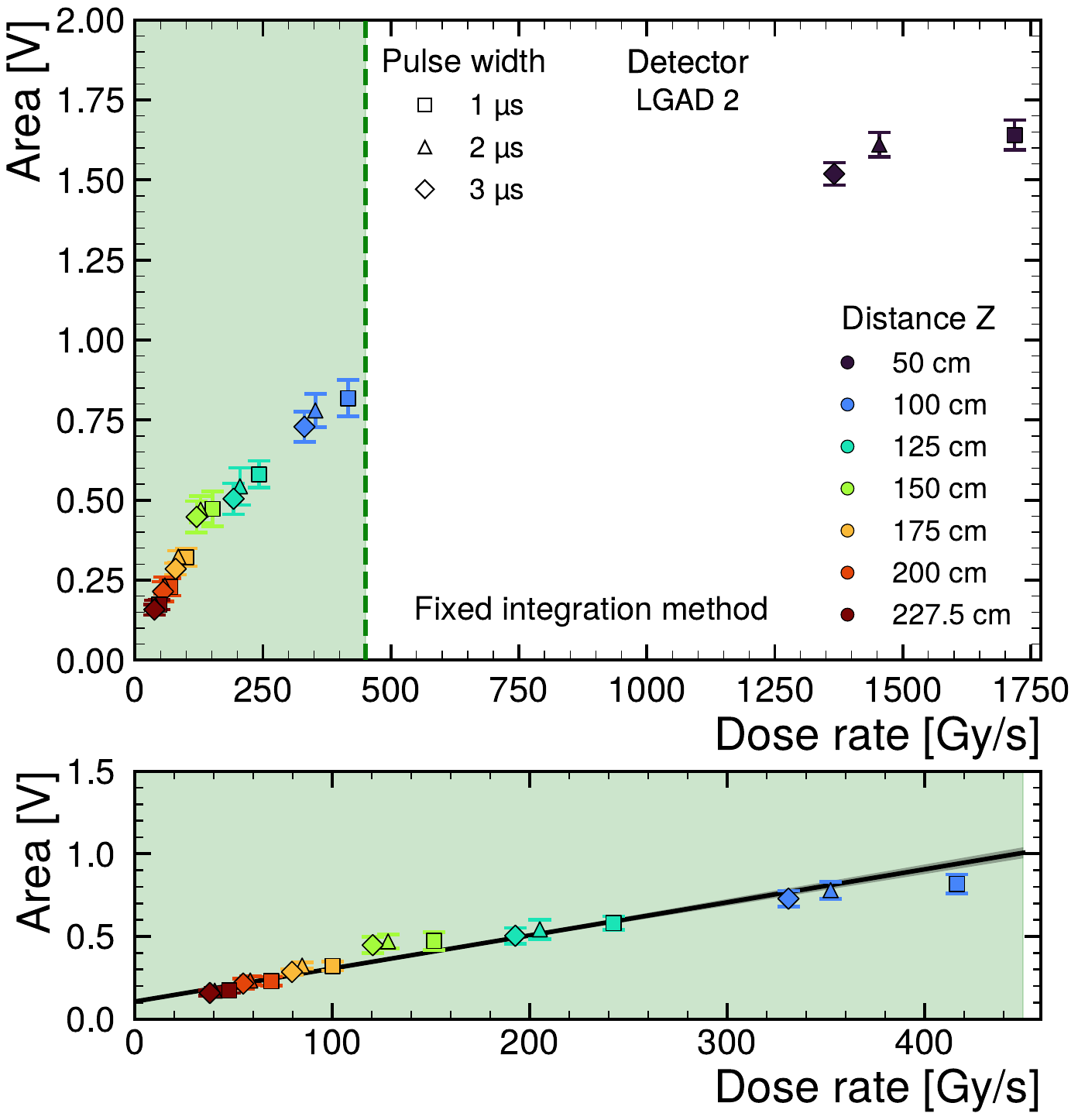}
    \hfill
    \includegraphics[width=0.47\textwidth]{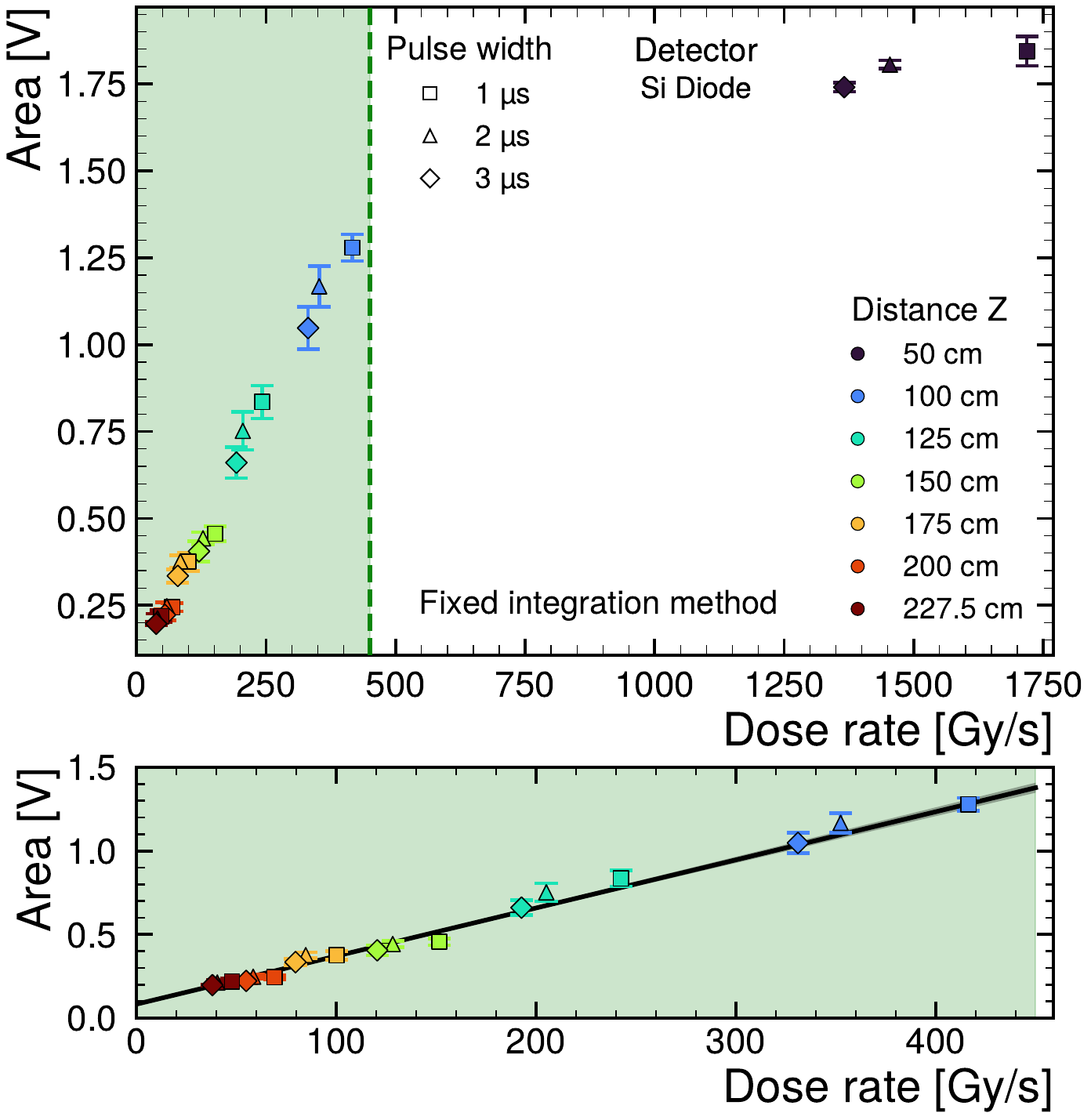}

    \vspace{0.35cm}
 
     \includegraphics[width=0.47\textwidth]{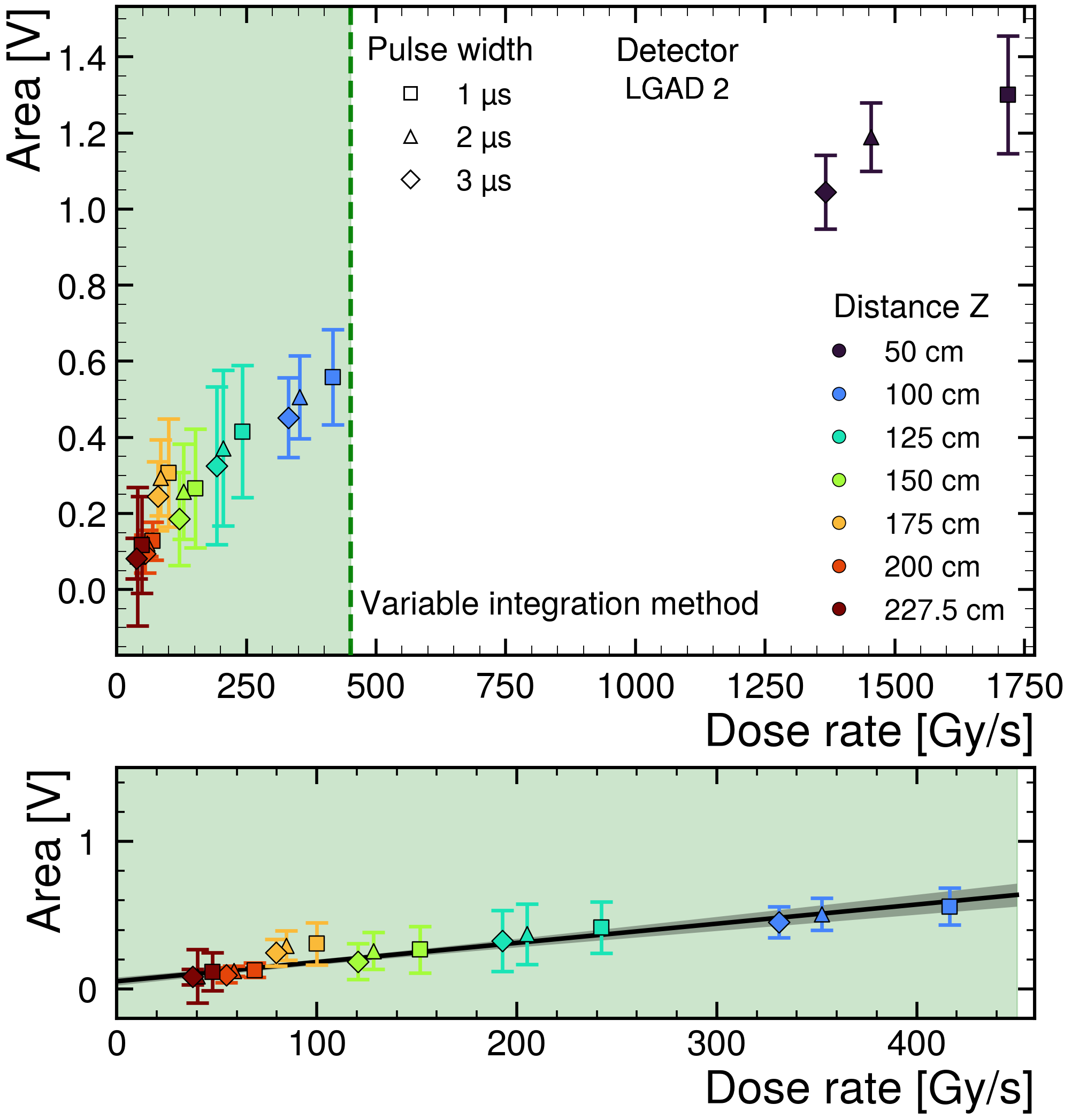}
    \hfill
    \includegraphics[width=0.47\textwidth]{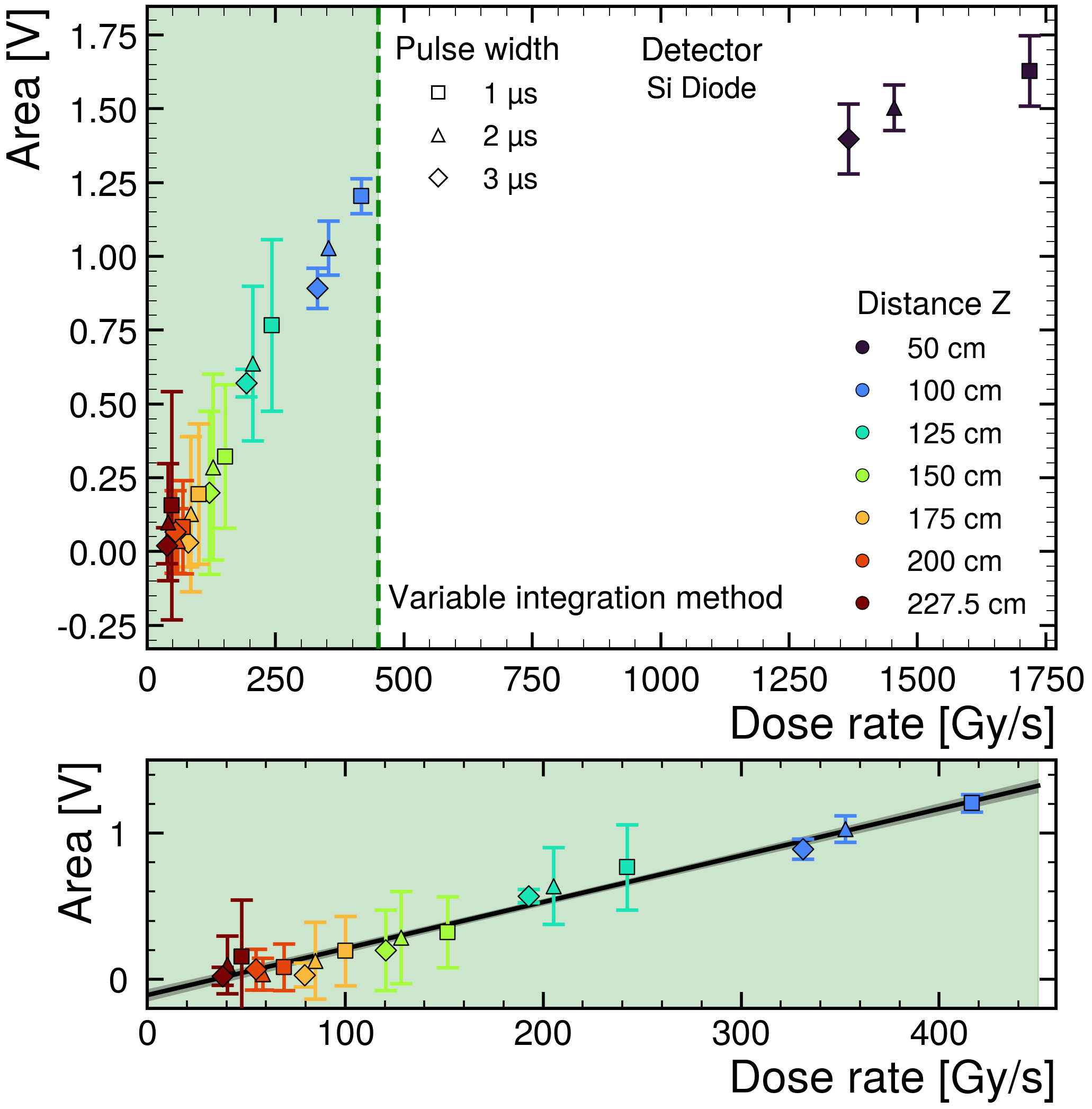}
\caption{
Charge versus dose rate (Gy/s)
at HV = 100~V for an electron 85V beam, combining all distances $z$ for LGAD 2 (left plots) and the Silicon diode (right plots) using the full integration method (upper plots) and the variable integration one (bottom plots). In green we show the zoomed linear region fit at low doses.
}
    \label{fig:charge_vs_dose_rate_allZ_HV100_BEAM85V}
\end{figure*}

Fig.~\ref{fig:charge_vs_dose_rate_allZ_HV100_BEAM85V} shows the correlation between the results of the dose measurement per second from the area method and the reference dosimeter by varying the distance from the beam exit. The charge versus dose rate (Gy/s)is displayed for LGAD 2 (left plots) and the Si diod (right plots) using 
using the full integration method (upper plots) and the variable integration one (bottom plots). In green we also show the zoomed linear region fit at low doses.

Each vertical value in Fig.~~\ref{fig:charge_vs_dose_rate_allZ_HV100_BEAM85V} corresponds to the average value from several pulses recorded at the same position and dose rate. Dose rates increase in the right direction in the horizontal axis, corresponding to closer distance to the beam exit (the different colors corresponding to different distances from the beam). The symbol shapes indicate the pulse widths going from 1 to 3 $\mu$s. In the top left plot, it can be noticed that the linear behaviour works up to $\sim$ about $\sim$450 Gy/s for $\mu$s pulses (so $\sim$ 4.5 10$^{-4}$ Gy/pulse). The advantage of our system is that we can measure the doses every $\mu$s or 100 ns, so almost instantaneously.  Beyond this point, the slope decreases (but the signal is still increasing) which is likely a sign of saturation due to higher doses at the proximity to the beam exit ($z\leq$125cm). 

The fact that the signal still increases up to $\sim$ 1800 Gy/s  (1.8 10$^{-3}$ Gy/pulse), even if the slope is modified (we do not see the signal flattening out). This means that we can still measure the dose up to these values once the detector is calibrated.  At the bottom of Fig.~\ref{fig:charge_vs_dose_rate_allZ_HV100_BEAM85V}, we display similar correlation results using the  variable integration range method and results are similar. 

The onset of the non-linear response at $\sim$$10^{-3}$~Gy/pulse points probably to a readout-chain limitation rather than a fundamental property of silicon. The limitation we observe is therefore not a property of the sensor material but of the readout chain, and in particular of the bandwidth and dynamic range of the LGAD~2 amplifier. This is an engineering constraint that is addressable with faster, higher dynamic range electronics, and we consider it the primary avenue for improvement in future work.

\subsection{Proton beams}

\begin{figure*}
    \centering
    \includegraphics[width=0.49\textwidth]{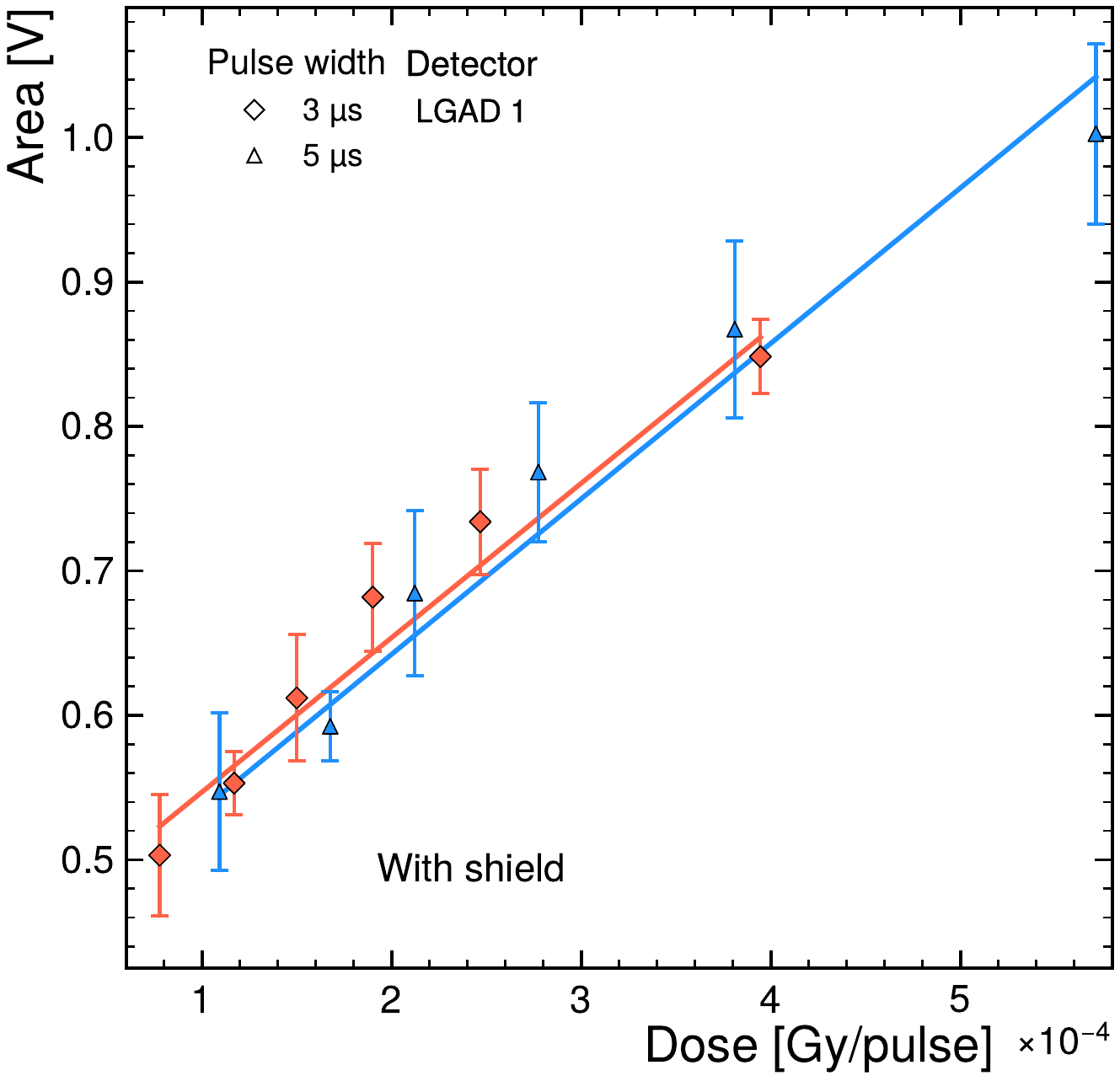}
    \hfill
    \includegraphics[width=0.452\textwidth]{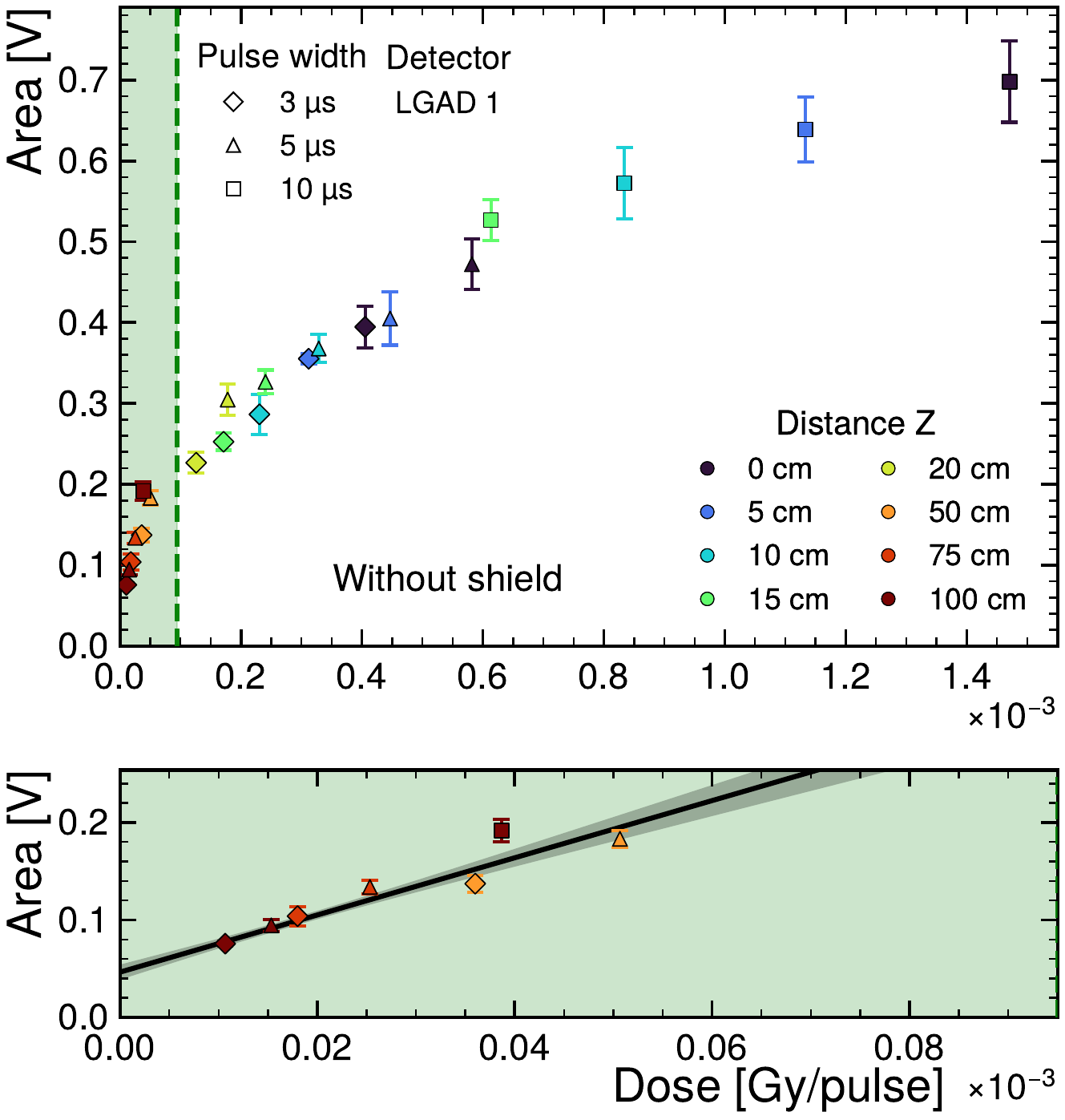}

    \caption{
    LGAD 1 response to proton beams as a function of dose rate (Gy/pulse) derived from the ion chamber measurements. Left:  Configuration using different widths of shielding for pulse widths of 3~$\mu$s and 5~$\mu$s. Right: Configuration without shielding but varying the distance from the beam exit for pulse widths of 3~$\mu$s, 5~$\mu$s, and 10~$\mu$s.}
    \label{fig:lgad1_with_without_shield_dose_rate}
    \label{fig11}
\end{figure*}

An initial measurement was performed by placing the detectors in front of the proton beam exit at the St.\ Louis Hospital, varying the size of a water filter (0--30~cm) and the pulse width (3~$\mu$s and 5~$\mu$s) to scan the delivered doses. We compare the instantaneous dose measurement from the ion chamber and from LGAD~1 as shown in Fig.~\ref{fig11}, left. A good correlation is observed up to $\sim 6\times10^{-4}$~Gy/pulse ($\sim$120~Gy/s), with the advantage that the measurement is performed instantanously.

An additional set of measurements was performed using the St. Louis Hospital proton machine. It involved varying the detector distance from the proton beam exit. The distances that were used are 0, 5, 10, 15, 20, 50, 75, 100 cm. In addition three different pulse widths were used namely 3, 5 and 10 $\mu$s. Fig~\ref{fig11}, right, shows the measurements  up to a dose of 1.5 10$^{-3}$ Gy/pulse, so about 150 Gy/s, consistent with the results shown in Fig.~\ref{fig11}, left. We notice a change of slope at low doses and the first linear region is highlighted in green. It is valid up to a dose of about 6. 10$^{-5}$ Gy/pulse, so about 12 Gy/s (please note that the left plot of Fig.~\ref{fig11} starts outside the green region of the right plot). These measurements correspond to the cases where the detector was further away from the beam exit ($z>$50cm). The sensors show non-linear response (compression) beyond this limit at higher doses with a slower increase in dose measurement with respect to the dosimeter results, at least up to about 200 Gy/s.

To summarize our studies with a proton beam, we see a linear behavior up to 12 Gy/s and a change of slope between 15 and 150 Gy/s. The non-linear response arises when multiple particles enter the detector within the same few-nanosecond window, so that the readout chain can no longer resolve them individually. As with the electron results, this is an engineering limitation of the current readout rather than a fundamental property of the silicon sensor. Using smaller detectors will allow increasing the maximum dose rate in the linear regime.

\subsection{Sensor active area scan}
\label{sec:scan}

The active area of the LGAD~1 sensor was scanned using a PiL106X picosecond laser available at KU, emitting 30~ps pulses at a wavelength of 1060~nm. The laser was operated at a frequency of 1~MHz with the attenuation varied from 10\% to 90\%. The active area was scanned in steps of 0.01~mm and an average of 10,000 pulses per scan position was recorded.

\begin{figure}[h!]
\centering
\includegraphics[height=0.7\textwidth]{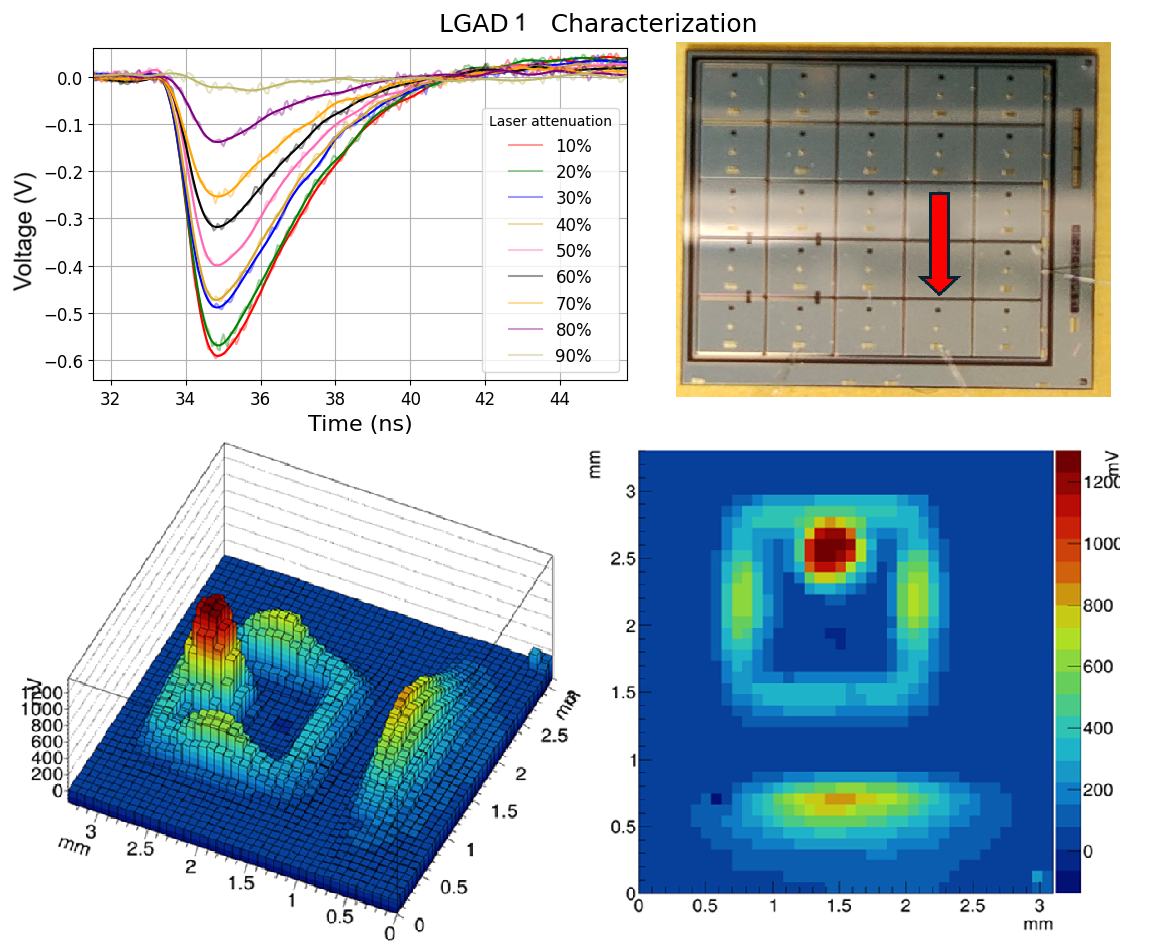}
\caption{\label{fig_scan1} Top left: LGAD~1 signal response to the picosecond laser source at a frequency of 1~MHz, for attenuation levels from 10\% to 90\%. Top right: Location of the scanned sensor (red arrow) within the full 25-sensor matrix. Bottom: Amplitude response mapped across the sensor active area, showing the non-uniformity that enables beam position determination.}
\end{figure}

Fig.~\ref{fig_scan1} shows the result of the laser scan of the LGAD~1 active area. The top left panel shows the sensor response as a function of laser attenuation, confirming a clean signal with a pulse duration of $\sim$7~ns consistent with the expected LGAD response. The top right panel identifies the scanned sensor within the full 25-sensor matrix. The bottom panel shows the amplitude response mapped across the active surface, revealing a non-uniform response. This non-uniformity allows a determination of the beam entrance position with a resolution of approximately 0.1~mm, demonstrating the potential for online beam position monitoring in future studies.

\section{Summary and conclusions}

We have demonstrated that Low Gain Avalanche Detectors (LGADs) and silicon diodes can measure instantaneous flash radiotherapy beam doses on microsecond timescales, using two complementary methods: a spike counting method and a signal area integration method. Measurements were performed with clinical electron and proton accelerators at the MD Anderson Cancer Center and the Siteman Cancer Center, covering dose rates from conventional to the FLASH-RT regime.

For electron beams (LGAD~2), a clear correlation between the integrated charge and the reference ion chamber dose was observed up to $\sim$1800~Gy/s using the area integration method and for proton beams (LGAD~1) up to $\sim$150~Gy/s, with a change of slope appearing respectively at 450 Gy/s and 12. Gy/s. The spike counting method is useful as a beam diagnostic tool at low dose rates (up to $\sim$1--2~Gy/s for our sensor sizes~\cite{Yepes2026}), confirming pulse structure and beam presence, but is not applicable for dosimetry at clinical or FLASH dose rates. The area integration method is the primary dosimetry tool, remaining effective when individual spikes overlap at higher doses.

Beyond the linear regime, the response continues to increase with a reduced slope and no true plateau is observed, showing the possibility to measure doses instantaneously after calibration.  %The comparison with SiC-based silicon devices, which have been shown to remain linear up to $\sim$4--5.5~MGy/s~\cite{Fleta2024, Milluzzo2024}, makes clear that 
The non-linearity we observe is likely not a fundamental property of silicon but reflects the bandwidth and dynamic range limitations of the current readout chain. This is an engineering constraint that can be addressed using higher-bandwidth amplifiers and smaller sensor active areas.

These results establish LGADs as viable candidates for online beam monitoring in flash radiotherapy. The per-pulse, microsecond-scale measurement capability demonstrated here is a prerequisite for real-time dose monitoring and beam interruption in clinical treatment. The identification of the readout chain as the current performance bottleneck provides a clear roadmap for future detector generations. The laser scan of the sensor active area further demonstrates a spatial resolution of $\sim$0.1~mm, pointing towards future beam position monitoring capability. Recommended next steps include tests with higher-bandwidth readout electronics, smaller active-area sensors, and a study of the effects of radiation damage and temperature variation on long-term LGAD performance.

\section*{Acknowledgements}

Christophe Royon, Tonatiuh Garcıa Chavez, Saul Anibal Rodrıguez
Ramırez, Harold Li thank NIH for its support under the contract number 1R21CA274193-01A1. Rachel Kovac-Fuentes performed part of this work supported by the National Science Foundation Graduate
Research Fellowship Program under Grant No(s) 1842494. Any opinions,
findings, and conclusions or recommendations expressed in this material are those of the
author(s) and do not necessarily reflect the views of the National Science Foundation.
Javier Murilo, Antonio Cota and Cristhian Cadena would also like to thank the Secretaría de Ciencia, Humanidades, Tecnología e Innovación (SECIHTI) in Mexico for providing support with the science at frontier project CBF-2025-I-2148 and Universidad de Sonora ACARUS area for providing supercomputing resources that were essential to conduct this research. \\

\section*{Bibliography}
\bibliographystyle{unsrt}
\bibliography{apssamp}

\end{document}